\documentclass{aastex}
\usepackage[onecolumn]{emulateapj5}
\usepackage{psfig}
\usepackage{amsmath}
\usepackage{euscript}

\newcommand{\cyg}{Cygnus X-3}

\slugcomment{Draft of \today}

\shorttitle{Flux ratios and light-travel time effects}
\shortauthors{Miller-Jones et al.\ }

\begin{document}

\twocolumn[ \title{Jet evolution, flux ratios and light-travel time effects}

\author{James C.\ A.\ Miller-Jones and Katherine M.\ Blundell}
\affil{University of Oxford, Astrophysics, Keble Road, Oxford, OX1
3RH, U.K.}
\and
\author{Peter Duffy}
\affil{Department of Mathematical Physics, University College Dublin,
  Dublin 4, Ireland}

\begin{abstract}
Studies of the knotty jets in both quasars and microquasars frequently
make use of the ratio of the intensities of corresponding knots on
opposite sides of the nucleus in order to infer the product of the
intrinsic jet speed ($\beta_{\rm jet}$) and the cosine of the angle
the jet-axis makes with the line-of-sight ($\cos{\theta}$), via the
formalism $I_{\rm a} / I_{\rm r} = \left((1 + \beta_{\rm jet}
\cos{\theta})/(1 - \beta_{\rm jet} \cos{\theta})\right)^{3+\alpha}$,
where $\alpha$ relates the intensity $I_{\rm \nu}$ as a function of
frequency $\nu$ as $I_{\rm \nu} \propto \nu^{-\alpha}$.  In the cases
where $\beta_{\rm jet} \cos{\theta}$ is determined independently, it
is found that the intensity ratio of a given pair of jet to
counter-jet knots is over-predicted by the above formalism compared
with the intensity ratio actually measured from radio images.  As an
example in the case of the microquasar Cygnus X-3 the original
formalism predicts an intensity ratio of $\sim 185$, whereas the
observed intensity ratio at one single epoch is $\sim 3$.
\citet{Mir99} have presented a refined approach to the original
formalism which involves measuring the intensity ratio of knots when
they are at equal angular separations from the nucleus.  This method
is however only applicable where there is sufficient time-sampling
(with sufficient physical resolution) of the fading of the jet-knots
so that interpolation of their intensities at equal distances from the
nucleus is possible.  This method can therefore be difficult to apply
to microquasars and is impossible to apply to quasars.  We demonstrate
that inclusion of two indisputable physical effects: (i) the
light-travel time between the knots and (ii) the simple evolution of
the knots themselves (e.g.\ via adiabatic expansion) reconciles this
over-prediction (in the case of Cygnus X-3 quoted above, an intensity
ratio of $\sim 3$ is predicted) and renders the original formalism
obsolete.
\end{abstract}

\keywords{methods: analytical --- radiation mechanisms: non-thermal
  --- relativity --- ISM: jets and outflows}
]
\section{Introduction}
\label{sec:intro}
Relativistic jets are observed in both quasars and microquasars, and
are often seen to consist of a series of discrete knots moving
outwards from a central nucleus, believed to correspond to the compact
object powering the outflow.  Measurements of the proper motions of
these knots are often used to constrain properties such as jet speeds
and inclination angles, and source distance
\citep[e.g.][]{Mir99,Hje95}.  The ratio of the intensities of
approaching and receding knots (if there is sufficient spatial
resolution that these can be accurately identified) have been used
\citep[e.g.][]{Sar97} to constrain their Lorentz factors, via
\begin{equation}
\frac{S_{\rm app}}{S_{\rm rec}} = \left(\frac{\mu_{\rm app}}{\mu_{\rm
    rec}}\right)^{k+\alpha} =
    \left(\frac{1+\beta\cos\theta}{1-\beta\cos\theta}\right)^{k+\alpha},
\label{eq:mirratio}
\end{equation}
where $\beta=v/c$ is the jet speed, $\theta$ is the inclination angle
of the jet axis to the line of sight, $\alpha$ is the spectral index
of the emission (defined by $S_{\nu} \propto \nu^{-\alpha}$, where
$S_{\nu}$ is the flux density at frequency $\nu$), $S_{\rm app}$ and
$S_{\rm rec}$ are the flux densities of a corresponding pair of
approaching and receding knots, $\mu_{\rm app}$ and $\mu_{\rm rec}$
are their proper motions, and $k=3$ for a jet composed of discrete
ejecta.

The luminosities $L(t)$ of the knots change with time $t$, as the
knots expand and the magnetic field, and hence the synchrotron
emissivity, decreases.  Thus the true flux ratio is
\begin{equation}
\frac{S_{\rm app}}{S_{\rm rec}} =
\left(\frac{1+\beta\cos\theta}{1-\beta\cos\theta}\right)^{k+\alpha}\frac{L_{\rm
    app}(t_{\rm app})}{L_{\rm rec}(t_{\rm rec})},
\label{eq:fluxratio}
\end{equation}
where $t_{\rm app}$ and $t_{\rm rec}$ are the times at which light
leaves the approaching and receding knots respectively in order to
arrive at the telescope at the same time.  Unless the jet axis is
perpendicular to the line of sight however, the light-travel time
between approaching and receding knots will mean that we see the
receding jet as it was at an earlier time, when it was more compact
and hence intrinsically brighter (but also dimmed in the observer's
frame by its recessional motion, taken into account by the original
formalism), compared with the approaching jet seen at the same
telescope time.  To account for this effect, \citet{Mir99} proposed
that the flux densities used to calculate the ratio should be measured
at equal angular separations from the nucleus.  This cannot always be
implemented in practice however, since this will require interpolation
unless good temporal coverage of the jets is available, or unless the
jet is a continuous flow, in which case the motion of individual knots
cannot be tracked in any case. At early times it may also be difficult
to separate the emission from moving jet knots and a fading core if
there is insufficient spatial resolution.  Moreover, as a result of
opacity or the presence of a broken power law, interpolation of the
spectrum may not be straightforward if in the observer's frame we
sample different parts of the spectrum at any given frequency.

In this \textit{Letter}, we present a method of using the flux ratios
from a single image of a source to constrain the jet speeds without
resorting to interpolation via the \citeauthor{Mir99} method.

\section{Flux ratios}
\subsection{Simple Scalings}

A synchrotron-emitting plasmon where the particles undergo adiabatic
expansion will have a power law decay in intensity, $L(t)\propto
t^{-\zeta}$, in which case equation\,\ref{eq:fluxratio} becomes
\begin{equation}
\frac{S_{\rm app}}{S_{\rm rec}} =
\left(\frac{1+\beta\cos\theta}{1-\beta\cos\theta}\right)^{k+\alpha}
\left(\frac{t_{\rm app}}{t_{\rm rec}}\right)^{-\zeta},
\label{eq:timeratio}
\end{equation}
and this scaling will apply to any process which gives a power law
decay in intensity.  We consider symmetric approaching and receding
jets, in which case after ejection at $t=0$, the epochs at which
photons leave corresponding points of the front and back plasmons,
$t_{\rm app}$ and $t_{\rm rec}$ respectively, are related by
\begin{equation}
\frac{t_{\rm app}}{t_{\rm rec}} =
\frac{1+\beta\cos\theta}{1-\beta\cos\theta}.
\label{eq:times}
\end{equation}
So in this simple case equation\,\ref{eq:timeratio} becomes
\begin{equation}
\frac{S_{\rm app}}{S_{\rm rec}} =
\left(\frac{1+\beta\cos\theta}{1-\beta\cos\theta}\right)^{k+\alpha-\zeta}.
\label{eq:powerratio}
\end{equation}

While this ratio is applicable to any process that gives a power law
decay, in the adiabatically expanding synchrotron case the parameters
$\alpha$ and $\zeta$ are not independent so that the determination of
the flux ratios by equation \,\ref{eq:powerratio} is actually not
introducing an extra parameter.


\subsection{Optically thin synchrotron emission and adiabatic
  expansion}
\label{sec:theory}

The total synchrotron emissivity from a single, optically thin jet knot
scales as \citep[e.g.][]{Lon94}

\begin{equation}
J(\nu) \propto B^{3/2}N(\gamma)\gamma^2\nu^{-1/2},
\end{equation}
where $B$ is the magnetic field strength, $\gamma$ is the Lorentz
factor of an individual electron assumed to be radiating at a single 
frequency

\begin{equation}
\nu = \left(\frac{\gamma^2eB}{2\pi m_{\rm e}}\right),
\end{equation}
and $N(\gamma)$ is the total number of electrons with energies in the
range $(\gamma$, $\gamma+d\,\gamma)$ in the plasmon,
given by
\begin{equation}
N(\gamma,t_0) = A\gamma^{-p},
\end{equation}
where $p$ is the electron index, $t_0$ is some arbitrary reference
time, and $A$ is the normalisation constant.  As the plasmon expands,
nonrelativistically, from a radius $R_0(t_0)$ to $R(t)$ the electron
energy scales as
\begin{equation}
\gamma = \frac{R_0}{R}\gamma_0,
\end{equation}
if synchrotron losses are negligible. The spectrum then 
evolves according to 
\begin{equation}
N(\gamma, t) =
\left(\frac{R}{R_0}\right)N\left(\frac{R}{R_0}\gamma_0,t_0\right).
\end{equation}
Putting all of the above together we find that the synchrotron
emissivity of an expanding plasmon is given by 
\begin{equation}
J(\nu)\propto\nu^{(1-p)/2}B^{(1+p)/2}R^{1-p}.
\end{equation}
As the plasmon expands the magnetic field strength will decrease and,
in the case of a tangled field, we have $B\propto R^{-1}$, so the
plasmon emissivity has a simple dependence on frequency and size given
by
\begin{equation}
J(\nu)\propto\nu^{(1-p)/2}R^{(1-3p)/2}.
\end{equation}
The ratio of flux densities as seen by the observer is then 
\begin{equation}
\frac{S_{\rm app}}{S_{\rm rec}} = \left(\frac{R(t_{\rm app})}{R(t_{\rm
    rec})}\right)^{(1-3p)/2}
    \left(\frac{1+\beta\cos\theta}{1-\beta\cos\theta}\right)^{k+(p-1)/2}.
\label{eq:observerratio}
\end{equation}
Although we could take the expansion of the plasmon to be of the form
$R\propto t^{\eta}$, it is particularly instructive to look at the
case of linear expansion, $\eta=1$, for which
equation\,\ref{eq:observerratio} becomes
\begin{equation}
\frac{S_{\rm app}}{S_{\rm
    rec}}=\left(\frac{1+\beta\cos\theta}{1-\beta\cos\theta}\right)^{k-p}.
\label{eq:ratio}
\end{equation}
This is the flux ratio observed at a given instant by the telescope as
opposed to the interpolated flux at equal angular separations. As a
simple generic case, emission from a jet composed of discrete ejecta
($k=3$) from a spectrum of electron index $p=2$ will give an exponent
of unity for equation\,\ref{eq:ratio}. By way of contrast, obtaining
an interpolated estimate at equal angular separations will, in this
case, give an exponent of $k+\alpha=3.5$.  The flux ratios to be
measured in the two cases would, however, differ, being measured in a
single image in the former case and at equal angular separations in
the latter.  A comparison of both methods in any given source would of
course be a useful means of inferring a possible asymmetry in the
approaching and receding jets \citep{Ato97}.

\subsection{Synchrotron Self-Absorption and Spectral Breaks}

When the particle spectrum contains a break or a turnover we must
adapt the above discussion. For example, in Cygnus X-3 (\citet{Mil04})
we observe two discrete knots, one on each side of the central
nucleus, with evidence for a turnover due to synchrotron
self-absorption. In this case the spectrum will have the form
$J_\nu\propto\nu^{5/2}$ below the turnover frequency $\nu_0$ when the
knot radius is $R_0$ while above this frequency the spectrum is
optically thin, $J_\nu\propto \nu^{-(p-1)/2}$. As the knot expands to
a radius $R$ its emission will then take the self-absorbed form

\begin{equation}
J(\nu,R)=
J_{\rm max}(R)
\left(\frac{\nu}{\nu_c(R)}\right)^{5/2},\;\;\;\nu\le\nu_c(R),
\end{equation}
while in the optically thin regime the intensity scales like
\begin{equation}
J(\nu,R)=J_{\rm max}(R)
\left(\frac{\nu}{\nu_c(R)}\right)^{-(p-1)/2},\;\;\;\nu\ge\nu_c(R).
\end{equation}

The critical frequency beyond which the emissivity becomes optically
thin is determined by 
\begin{equation}
\nu_c(R)=\nu_0\left(\frac{R}{R_0}\right)^{-(3p+4)/(p+4)},
\end{equation}
and the emission at that frequency, which is the peak of the knot
spectrum, becomes
\begin{equation}
J_{\rm max}(R)=J_0\left(\frac{R}{R_0}\right)^{-5p/(p+4)}.
\end{equation}

Turning now to the flux ratios observed at a given instant and
frequency it is clear that at sufficiently early and late times we
will have two extremes. In the former case, when the observed emission
from each knot is optically thick we will see a $J_\nu\propto
R^{5/2}\nu^{5/2}$ spectrum from each and the flux ratio exponent is
$k=3$ for discrete ejecta. However, it may be difficult to observe
actual knots in this regime without mixing in possible nuclear
emission. At observed frequency $\nu$ the emission will become
optically thin from the approaching knot when its radius is $R_1$
which is determined by 
\begin{equation}
\nu=(1+\beta\cos\theta)\nu_0\left(\frac{R_1}{R_0}\right)^{-(3p+4)/(p+4)},
\end{equation}
while the emission from the receding knot will remain optically thick,
at this frequency, 
until the approaching knot has a radius of $R_2$ which can be easily
shown to be 
\begin{equation}
R_2=\left(\frac{1 +
\beta\cos\theta}{1-\beta\cos\theta}\right)^{2p/(3p+4)}
R_1.
\end{equation}

Therefore, when the approaching knot has a radius $R$ satisfying
$R_1\le R\le R_2$ the observed emission at frequency $\nu$ will be a
mix of Doppler boosted, optically thin emission from the forward knot
and optically thick emission from the receding component. The flux
ratio in this regime is now dependent on time, i.e. knot radius, and is
given by
\begin{equation}
\frac{S_{\rm app}}{S_{\rm
    rec}}=\left(\frac{1+\beta\cos\theta}{1-\beta\cos\theta}\right)^{k}
\left(\frac{R}{R_1}\right)^{-(3p+4)/2}.
\label{eq:mixratio}
\end{equation}

At $R=R_2$ all of the emission becomes optically thin at this
frequency and the flux ratios predicted by equations \ref{eq:ratio}
and \ref{eq:mixratio} are equal. Therefore, the flux ratio exponent
drops from a value of $k$ to $k-p$ as the front and then the receding
knot become optically thin. During this time the spectrum at
frequency $\nu$ also evolves from $\nu^{5/2}$ to $\nu^{-\alpha}$ and
the forward knot expands by a factor $R_2/R_1$, where both radii are
frequency dependent. The time taken for this expansion is determined
by the expansion velocity $V$ of the knot.

In reality however, it is unlikely that the turnover in the spectrum
would occur at one single frequency.  There would be a finite turnover
region in which the spectrum evolved from a $\nu^{5/2}$ power-law to
$\nu^{-(p-1)/2}$.  Depending on the width of the turnover region and
the value of $\beta\cos\theta$, the receding knot could be in the
turnover region of the spectrum by the time the approaching knot had
become optically thin.  In this case, the above results would not be
strictly applicable.

Nonetheless, these general points apply to any source of opacity which
changes the spectral shape or indeed to any broken power law that
might be attributable to the acceleration mechanism. It presents the
possibility that the evolution of the spectrum from flares in
microquasars may well be influenced by the light-travel time
differences between approaching and receding knots, as outlined in
\citet{Mil04}.

\subsection{Caveats}

Care should be taken if the expansion mode of the plasmons changes
prior to the observation from which the flux ratio is derived.  In
such a case, for example a transition from slowed to free expansion
\citep{Hje88}, the time decay of the flux density would change
(steepen with time in this case).  In order to use flux ratios to
constrain the value of $\beta\cos\theta$, the flux densities of the
approaching and receding knots would then have to be measured when the
knots were both in the same expansion regime.  Unless the transition
radius were known, this would require actually measuring (as opposed
to interpolating) the flux densities at equal angular separation from
the core.  We note that if there is significant deceleration of the expanding
plasmons due to interaction with surrounding material, as mentioned by
\citet{Hje95a}, then $(R/R_0)\propto(t/t_0)^\eta$, where $\eta<1$, and
equation \ref{eq:ratio} then requires modification.  We also draw
attention to \citet{Fen03}, which presents caveats to be considered
when using proper motions to place limits on the bulk Lorentz factors
of jets; any Lorentz factors thus derived are strictly only
\emph{lower limits}.

\section{Comparison with observations}
The VLBA observations of \cyg\ presented by \citet{Mil04} show a jet
which at 5\,GHz and 15\,GHz appears to be composed of two separating
discrete knots, which were interpreted as approaching and receding
plasmons.  A precession modelling analysis yielded a value
$\beta\cos\theta = 0.62\pm0.11$, and the spectral index of the
emission was found to be $\alpha=0.60\pm0.05$.  Assuming linear
expansion of the jet knots, we would thus predict a flux density ratio
of $3.2\pm1.0$.  For the last two epochs (2001 September 20 and 21),
the measured flux ratios are given in Table\,\ref{tab:cyg}.  While not
matching the theoretical prediction perfectly, they are now of the
correct order, in contrast with the predictions of the original
formalism, which is wrong by two orders of magnitude.  There are
various possible explanations for the slight discrepancy.  Most
importantly, the measurement of the flux densities themselves was
often difficult.  It is also possible that the plasmon expansion was
not exactly linear with time, which would alter the exponent in
equation\,\ref{eq:ratio} and change the predicted flux density ratio.
Moreover, the measured spectral index $\alpha$ was for the integrated
spectrum; the values of $\alpha$ and $p$ could in principle differ for
the individual jet knots.  The quality of the data makes it difficult
to interpolate back to the flux densities at equal angular separations
in this case, but our best attempts gave flux ratios between 1.58 and
10.63.  In such cases, our direct measurement method gives a much more
accurate determination of the expected flux ratio, for more meaningful
comparison with the jet speeds and inclination angle found by
different methods.  We note that if the spectral index of the jet
material is known, our method requires only a single image to
determine the value of $\beta\cos\theta$, whereas the interpolation
method requires at least two images taken at different times.  This
frees it from the uncertainty inherent in comparing VLBI images,
particularly if the imaging is difficult, as was the case in these
observations \citep{Mil04}.  For a single image, the ratio of two flux
densities is set, whereas when comparing different images, in order to
be able to interpolate accurately, one has to be confident that one
has recovered the same fraction of the true flux density in both
images in order to be able to take an accurate flux density ratio.

This theory could also be applied to the observations of GRS\,1915+105
detailed by \citet{Mir94}.  They observed discrete radio ejecta moving
outward from the nucleus over a period of $\sim 1$ month.  Again, we
take the flux density ratio of their observed knots once they had
clearly separated from one another and from the nucleus, and we only
compare corresponding pairs of ejecta.  From their derived value of
$\beta\cos\theta = 0.323\pm0.016$ and their quoted spectral index of
$\alpha=0.84\pm0.03$, we predict a flux density ratio of
$1.24\pm0.05$.  For the later epochs (1994 April 16, 23 and 30), the
measured flux density ratios are 2.33, 2.63 and 1.80 respectively.
Again, this is slightly greater than we predict, but is of the right
order.  Underpredicting the flux density ratio implies the exponent
should be larger in equation\,\ref{eq:ratio}, which requires $\eta>1$,
i.e.\ the expansion scales slightly more rapidly with time than
$R\propto t$.

\section{Conclusions}
We have considered the evolution of synchrotron bubbles (plasmons) in
oppositely-directed microquasar jets.  We have found that our new
formalism can explain the observed flux density ratios in microquasar
jets in systems in which the synchrotron bubble model is applicable,
such as Cygnus X-3.  In contrast, the original formalism considerably
overpredicts the observed flux density ratio in observations of this
system.  In the case of free (linear) expansion,
$(R/R_0)\propto(t/t_0)$, we found that the flux ratios of the
approaching and receding plasmons are given by $S_{\rm app}/S_{\rm
rec} = ((1+\beta\cos\theta)/(1-\beta\cos\theta))^{k-p}$.

\acknowledgments
J.C.A.M.-J. thanks the UK Particle Physics and Astronomy Research
Council for a Studentship.  K.M.B.\ thanks the Royal Society for a
University Research Fellowship.  K.M.B.\ and P.D.\ acknowledge a joint
British Council/Enterprise Ireland exchange grant.

\clearpage
\onecolumn

\begin{deluxetable}{ccc}
\tabletypesize{\scriptsize}
\tablewidth{0pt}
\tablecolumns{3}
\tablecaption{\cyg\ VLBA flux density ratios measured from
  observations of 2001 September outburst of \cyg\ \label{tab:cyg}}
\tablehead{
\colhead{Date (UT)} & \colhead{Observing Frequency (GHz)} &
\colhead{Flux density ratio (South/North)}
}
\startdata
September 20 & 5 & $1.43\pm0.05$ \\
September 21 & 5 & $2.39\pm0.10$ \\
September 20 & 15 & $1.14\pm0.19$ \\
September 21 & 15 & $3.09\pm0.14$ \\
\enddata
\end{deluxetable}

\end{document}